# Stepping out of Flatland: Discovering Behavior Patterns as Topological Structures in Cyber Hypergraphs

Helen Jenne, Sinan G. Aksoy, Daniel Best, Alyson Bittner, Gregory Henselman-Petrusek, Cliff Joslyn, Bill Kay, Audun Myers, Garret Seppala, Jackson Warley, Stephen J. Young, Emilie Purvine

Data breaches and ransomware attacks occur so often that they have become part of our daily news cycle. Last year, 1,802 data compromises affected 422 million people [31]. In a 2022 op-ed co-written by the CISA Director and National Cyber Director, they described the omnipresent threat of cyber attacks as "the new normal," writing that in the modern landscape of complex cyber threats, "our shields will likely be up for the foreseeable future" [12]. This is due to a myriad of factors, including the increasing number of internet-of-things devices, shift to remote work during the pandemic, and advancement in adversarial techniques, which all contribute to the increase in both the complexity of data captured and the challenge of protecting our networks. At the same time, cyber research has made strides, leveraging advances in machine learning and natural language processing to focus on identifying sophisticated attacks that are known to evade conventional measures. While successful, the shortcomings of these methods, particularly the lack of interpretability, are inherent and difficult to overcome. Consequently, there is an ever-increasing need to develop new tools for analyzing cyber data to enable more effective attack detection.

In this paper, we present a novel framework based in the theory of hypergraphs and topology to understand data from cyber networks through *topological signatures*, which are both flexible and can be traced back to the log data. While our approach's mathematical grounding requires some technical development, this pays off in interpretability, which we will demonstrate with concrete examples in a large-scale cyber network dataset. These examples are an introduction to the broader possibilities that lie ahead; our goal is to demonstrate the value of applying methods from the burgeoning fields of hypernetwork science and applied topology to understand relationships among behaviors in cyber data.

## From Flatland to Cyberland

In U.S. and NATO military doctrine, cyberspace is often considered as a domain, alongside land, sea, air, and space [13, 23]. But unlike these domains we cannot travel into cyber space and physically touch the system processes or digital information. Because of the lack of physical constraints in cyber, many have observed that thinking of it as a domain could be a hinderance by putting unnecessarily restrictive limits on what behaviors are possible [23, 34, 36]. In a recent article [36], Pierre Trepagnier of MIT Lincoln

Laboratory contended that instead we should think of cyberspace as another dimension. He presented his argument using the classic novella *Flatland*, by Edwin A. Abbott, which takes place in a two-dimensional world inhabited by geometric figures. In the story, the narrator, a square, is visited by a sphere from three-dimensional Spaceland, who appears to be a circle as it passes through Flatland. Despite the sphere's best efforts, the square cannot comprehend three-dimensional space until the sphere takes him there. In his essay, Trepagnier identifies numerous parallels between our challenges to understand cyber space and the square's inability to understand Spaceland. He writes,

> *In our physical world, our senses have evolved to perceive threats directly. But we cannot perceive packets; our perception of cyber is entirely synthetic, through sensors which we place out in Cyberland and whose outputs we route back into [our world].*

To elaborate on Trepagnier's description, sensors in Cyberland collect timestamped data called logs. Log data are generated from every device and application in the network, including routers, firewalls, workstations, and servers. Logs consist of digital observations, which vary in terms of the granularity, scope, and modality of information captured. For example, host logs are records of everything that happens at a system level, where a host is an individual computer, server, or other connected device. These records include login attempts, file creation and deletion, registry edits, errors, warnings, and other processes. Network logs offer a higher level perspective; they record information that crosses the boundaries of individual hosts like the flow of digital packets between hosts, and thus offer valuable insights into network-wide patterns. Gaining a complete picture of the current cyber landscape requires synthesizing information from host logs, network logs, and other data sources. Each stream of logs provides incomplete information, but they are linked through the imperceptible dimension of Cyberland. The challenge is modeling these data in such a way that recovers the unseen complexity and allows cyber analysts to take action based on the understanding they gain from the models.

## *OpTC data: Introduction and context*

Before describing our approach to modeling cyber data, we introduce the dataset we will use throughout this article: the Operationally Transparent Cyber (OpTC) dataset [11]. The OpTC dataset was released by the Defense Advanced Research Projects Agency (DARPA) Transparent Computing program to enable research that enhances understanding of and defense against advanced persistent threats (APTs) at scale. An attack by an APT is an extended cyber attack (low and slow) that, for example, aims to steal highly sensitive data or to impose the will of the malicious actor. It involves multiple stages, including initial infiltration, expanding access, using that access to gain administrative credentials, moving laterally to other servers or workstations, and finally exfiltrating data.

The OpTC dataset contains over 17 billion events generated from a simulated network consisting of approximately 500 hosts, scripted to mimic daily user activities (e.g., downloading files from emails, editing files, and browsing the internet), along with three days of annotated red team activity representing APT scenarios. The OpTC dataset brings both network and host logs together in a common format with common metadata fields, allowing one to make connections between system processes and other individual log events. We show an excerpt of the logs in Figure 1 to illustrate some of the detail provided in the dataset. These are a particular type of network logs, called network flow logs, that summarize host-to-host communications, such as a computer accessing a website or a user opening a remote desktop connection to access another computer. In the OpTC data, a single network flow log contains the start time of the flow (timestamp), the length of the flow (duration), the originating computer system (source IP), the recipient (destination IP), the interface used to send and receive communication (source and destination port), and the method of sending data (protocol), in addition to other metadata such as the path to the program that started the event (image path) and which entity is performing the action (principal). The richness of the data, documentation of red team attacks in the ground truth, and subsequent in-depth analysis of the network and host events [4] has led to the use of OpTC in a variety of research programs for cyber security threat detection [5, 7, 10, 16, 26].

| time | action-object | host | principal | pid | source IP | dest IP | dest port | protocol | image path |
|---|---|---|---|---|---|---|---|---|---|
| 9/24 13:51:28 | START-FLOW | SysClient0501 | bantonio | 2956 | 142.20.57.246 | 142.20.61.189 | 3389 | TCP | mstsc.exe |
| 9/24 13:52:08 | MESSAGE-FLOW | SysClient0974 | sbobertz | 3768 | 142.20.59.207 | 153.129.45.5 | 80 | TCP | firefox.exe |
| 9/24 13:54:36 | MESSAGE-FLOW | SysClient0974 | sysadmin | 3636 | 142.20.59.207 | 142.20.56.6 | 3389 | TCP | mstsc.exe |
| 9/24 13:55:40 | START-FLOW | SysClient0811 | rsantilli | 5712 | 142.20.59.44 | 142.20.61.130 | 135 | TCP | python.exe |

**FIGURE 1.** An excerpt of the network flow logs of the OpTC data. These are only a fraction of the keys (columns) recorded in the data. There are common keys shared among all log types, like action-object, host, principal, process ID, and others, but also fields unique to each log type. [**FIGURE 1 ALT TEXT.** A table of network flow records.]

## Hypergraphs as models of cyber network data

Activity in the OpTC data (and cyber log data in general) is characterized by many complex relationships among groups of items recorded in the logs. For example, groups of ports can be related by virtue of the processes that use them; or groups of IPs can be related based on the protocols they employ for their various communications. To model and analyze the relationships present in complex data like the OpTC data, we turn to hypergraphs. As mathematical models of data with group-wise relationships, hypergraphs have provided great benefit in recent years to researchers across a variety of domains [17, 19, 22, 24, 27, 30].

A hypergraph consists of a set of vertices, representing individual entities, along with a collection of hyperedges, where each hyperedge is a subset of the vertices of any size and represents some joint property among the vertices it contains. For example, the relationship between users and destination ports can be naturally structured as a hypergraph where the users are vertices, and each destination port hyperedge contains the users that use that port, as shown in Figure 2 (left and right). In general, when data come as individual records and each record contains the same field names, one can choose two fields (e.g., source IP and destination port in network flow logs, or command line and user in host logs) and model the relationship among the items that fill those fields using a hypergraph.

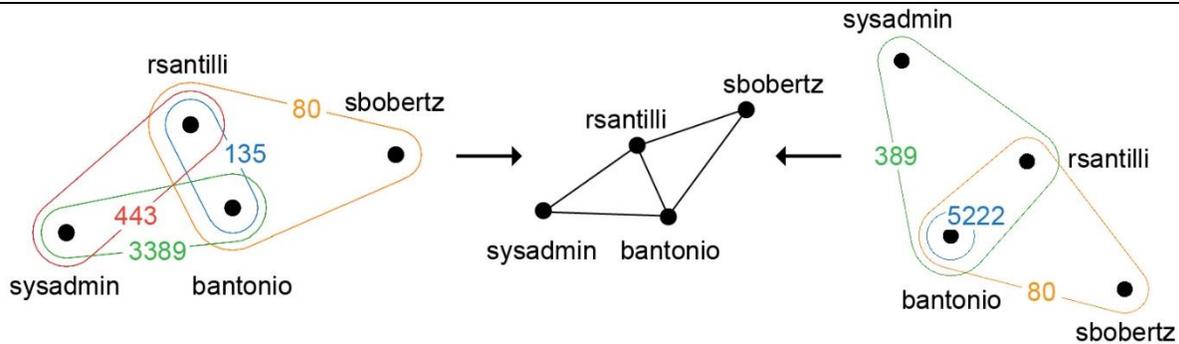

**FIGURE 2.** Two user–port hypergraphs (left and right) that have the same graph representation (center). Both hypergraphs capture user behavior information through the hyperedge containment and intersection patterns that the graph cannot. [**FIGURE 2 ALT TEXT.** Two hypergraphs and one graph.]

As the name suggests, hypergraphs are generalizations of graphs, which model pairwise relationships by restricting hyperedges (called "edges" in this context) to be size two. In cyber data, graphs have been used to model communications between IP addresses, relationships between users, and similarities among malware [2, 3, 6, 7, 9, 20, 33]. But graphs should be used with caution in data with complex relationships; as we learned from Flatland, appearances when projecting down from higher dimensional spaces can be deceiving. For example, the user–port relationship described above with a hypergraph can also be represented as a graph where two user vertices are linked by an edge when they perform actions using the same port. However, modeling the data with a graph results in information loss: the same user–user graph can correspond to many user–port hypergraphs, as shown in Figure 2. As the examples in this figure illustrate, hypergraphs can capture complexities that graphs cannot. While two edges in a graph can only interact by sharing one vertex, two hyperedges in a hypergraph can interact in many more ways. Hyperedges can intersect in any number of vertices, or one hyperedge can be fully contained within another. If we think of hyperedges as representing groups of vertices that share a behavior, the hypergraph perspective allows us to model how behaviors are linked or related by virtue of the vertices that display that behavior. The question then is how to use these models to gain understanding of the data and the behaviors therein.

A recent and active area of research to study hypergraph models of data is "hypernetwork science" [1], taking inspiration from the well-established field of "network science,"[1] which is the analysis of graph models of real-world data. Often in hypernetwork science, methods that were developed to analyze graph models are extended and generalized to apply to hypergraphs; but in many cases this is not a trivial task. One such graph method that has proven valuable across a wide range of fields is the study of how small, connected graph substructures (also known as "motifs") appear in large graphs. A graph motif is nothing more than a small connectivity pattern or signature. Some of the simplest graph motifs include the triangle, square, n-star, and n-path. All 3-edge motifs—the triangle, 3-star, and 3-path—are illustrated in Figure 3 (left). In cyber graphs, the "connector," where multiple vertices all connect to the same pair of vertices (an example with three central vertices is pictured in Figure 3), and n-star motifs are common and indicative of certain kinds of network behaviors and configurations [35].

---

[1] The terms "graph" and "network" are often used synonymously, with "network" usually connoting when a graph is modeling real-world data. However, from here on we will use the term "graph" to avoid confusion with the cyber use of the word "network."

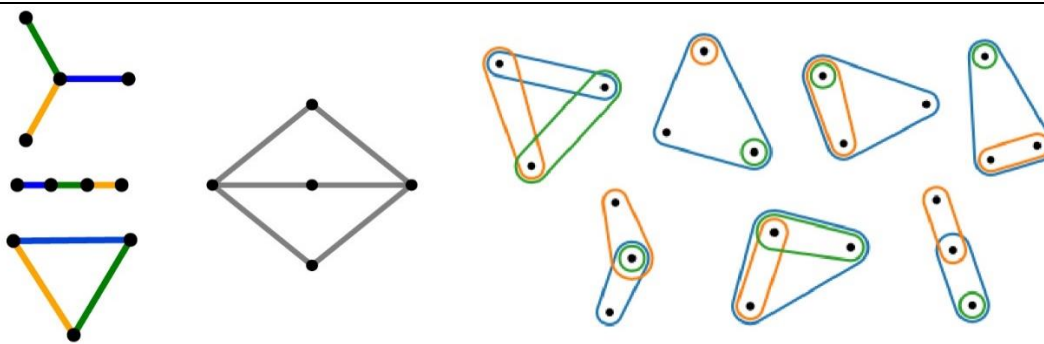

**FIGURE 3.** Three-edge graph motifs (first column), an example of a connector motif (second column), and seven of the 27 three-hyperedge hypergraph motifs (right). [**FIGURE 3 ALT TEXT.** Four graph motifs and seven hypergraph motifs.]

In order to extend motifs to hypergraphs, with the straightforward generalization being small, connected subhypergraphs, we look to recent work that defines 27 hypergraph motifs with three hyperedges [25] and develops algorithms to count them. But the authors of that work note that generalizing their hypergraph motifs to four and five hyperedges results in a combinatorial explosion: there are 1,853 four-hyperedge hypergraph motifs (compared to 5 four-edge graph motifs) and over 18.6 million five-hyperedge hypergraph motifs (compared to 12 five-edge graph motifs). To illustrate just some of the complexities of hypergraph motifs, we again refer to Figure 3 (right) where seven of the 27 three-hyperedge hypergraph motifs are shown, those that can occur with just three vertices. As we will see in the OpTC data, these complexities, in particular the presence of single-vertex hyperedges and hyperedge containment, are important for capturing different behavior patterns. Because of their combinatorial complexity, for hypergraph motifs to be useful in characterizing these patterns, they must be studied as a small number of groups of motifs where each group contains motifs of "similar enough" structure. While one could imagine many ways to define similar enough structure, resulting in different groupings of motifs, one way is to turn to the mathematical field of topology, which provides techniques for summarizing structure in a flexible and interpretable way. *This forms the basis of our framework; we define topological signatures in hypergraphs by grouping motifs that are topologically similar.*

## The topological perspective

Topology studies different shapes with a flexible notion of what it means for two shapes to be "the same", where two shapes are considered the same if they can be continuously deformed (via stretching, twisting, bending, but not breaking, tearing, or puncturing) into one another. Figure 4 shows a classic example of a continuous deformation between a coffee cup and a donut. This concept is made rigorous by studying what are called topological invariants. A topological invariant is a mathematical quantity that identifies certain properties of shapes and is guaranteed to be equal for two shapes that are the same topologically. For the last 20 years, topological invariants have been used as a tool to make sense of high-dimensional data in a field called topological data analysis. One of the most studied topological invariants in the data science community is called homology, which informally is a method of counting holes in a shape. The number of connected components (0-dimensional holes), loops (1-dimensional holes), voids (2-dimensional holes, like the inside of a basketball), and higher dimensional analogs are topological invariants that conveniently encode multi-dimensional structures in a way that has proven useful in many applications and for different data types [8, 15].

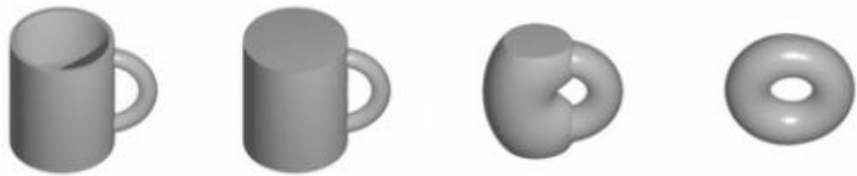

**FIGURE 4.** A donut and a coffee cup are topologically the same because the coffee cup can be continuously deformed into a donut. Image credit: Lucas Vieira, Public domain, via Wikimedia commons. [**FIGURE 4 ALT TEXT.** Four stages of deforming a coffee cup into a donut.]

Core to the notion of homology is the idea of a continuous deformation. But it is not immediately clear what that means in the case of a hypergraph. While we can draw a hypergraph on a page, it is not something we can hold in our hands and twist or bend. To leverage the topological invariant of homology to create topological signatures in the context of hypergraphs, we need a method to translate hypergraphs into topological objects that we can deform. Example building blocks of such tangible topological objects, in dimensions we can perceive, are points (0-dimensional), lines (1-dimensional), solid triangles (2-dimensional), and solid tetrahedra (3-dimensional); these building blocks are shown connected together to form an example topological object in Figure 5. It will be important later to know that these building blocks also include their lower dimensional boundaries. For example, a triangle includes its three edges and three vertices; a tetrahedron includes its four triangular faces, six edges, and four vertices. Higher dimensional analogs exist, and even though we cannot hold them, the theory for computing their homology extends seamlessly. Translating from a hypergraph to one of these topological objects can be done in many ways, and the topological holes of each one will tell us something different about the hypergraph and the data that was used to construct it.

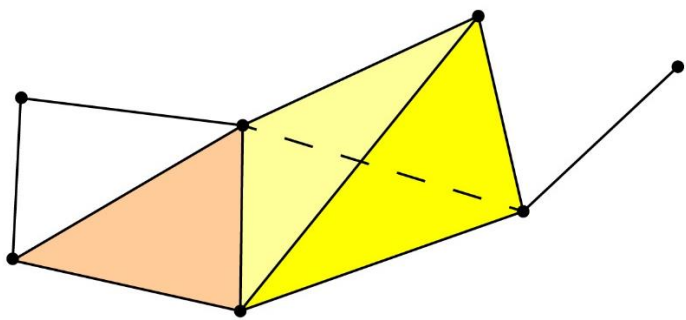

**FIGURE 5.** A topological object composed of points, lines, solid triangles (orange), and solid tetrahedra (yellow). [**FIGURE 5 ALT TEXT.** Points, lines, a triangle, and a tetrahedron joined together.]

*Translating hypergraphs into the language of topology*

Our research team is exploring various methods to translate hypergraphs into topological objects and what insight the homology of these objects provides in the context of cyber data. As we will see, after finding a hole in a topological translation of a hypergraph, we can identify the hypergraph motif that gives rise to the hole. When multiple hypergraph motifs give rise to the same dimension of hole, we group them as motifs that are topologically similar through the lens of that translation. This group of motifs is then a topological signature.

We have identified two hypergraph translation methods of particular interest, which result in two types of groupings of motifs into topological signatures, but there are others that could also lead to interesting insights into cyber datasets [18, 32]. We briefly mention the first and provide a reference to work where we show its use in cyber. We then go into more detail in the second construction, including grounding it by interpreting the holes in the context of benign and adversarial activity in the OpTC data.

Perhaps the most straightforward way to create a topological object from a hypergraph is by replacing each hyperedge of size k with a solid *(k-1)-*dimensional building block. We will refer to this as the closure of the hypergraph. In Figure 6, we show three example hypergraphs (first column) with their closures (second column). Notice that two of the hypergraphs in this figure have the same closure, which has a 1-dimensional hole. Therefore, these hypergraphs are topologically similar motifs through the lens of the hypergraph closure. This is because forming the closure of a hypergraph is akin to adding every subset of every hyperedge to the hypergraph, whether or not it was there to begin with. When a hyperedge of size four is present, for instance, we replace it with the solid tetrahedron that also includes all the 3-way triangular faces, 2-way edges, and singleton vertices. Recently our team studied how motifs giving rise to holes in the closure of hypergraphs that capture the relationship between source IP and executable (e.g., python.exe) in small time windows in the OpTC data change over time [28]. In that work unusual patterns of topological change correlated with some instances of malicious network behavior recorded in the ground truth.

Motifs found using the closure of a hypergraph identify what we might consider as intrinsic cycles, those that you can see if you just look at a drawing of the hypergraph. This is certainly a valuable perspective but forming the closure only provides one way of studying how behaviors, captured by hyperedges, interact. Another approach comes from the earlier observation that a hyperedge can be interpreted as a group of vertices sharing a behavior. From this perspective we posit that hyperedge containment relations are important since they mean that a group of vertices that share one behavior (the smaller hyperedge) also fully share another behavior (the larger hyperedge). This position leads us to introduce our second method to translate a hypergraph into a topological object that we call the nesting object of the hypergraph. To put it concisely, the nesting object includes a vertex for each hyperedge in the hypergraph and a *(k-1)-*dimensional building block for every sequence of k hyperedges where one is a subset of the next. To explain how this construction relates to our assertion that hyperedge containments are important, we build the nesting object in a two-step process.

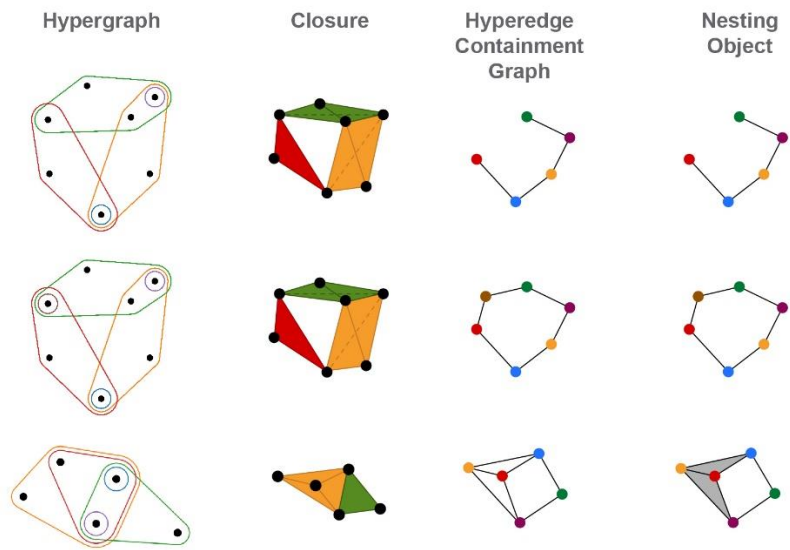

**FIGURE 6:** Three small hypergraphs with their closure, hyperedge containment graph, and nesting object. In the closure, the topological building blocks (triangles and tetrahedra) are colored corresponding to the hyperedges that give rise to them. In the hyperedge containment graph and nesting object, the vertices are colored the same as their corresponding hyperedges. [**FIGURE 6 ALT TEXT.** A table of hypergraph visualizations, their closures, hyperedge containment graphs, and nesting objects.]

In the first step of the translation, we create the hyperedge containment graph (HCG) of a hypergraph that has one vertex for each hyperedge, and an edge in the HCG represents a hyperedge containment relationship in the hypergraph. Concretely we add an edge in the HCG between the vertices corresponding to hyperedges *e* and *f* whenever *e* is contained in *f*, or vice versa. The third column of Figure 6 shows the HCG for the same three example hypergraphs. Since any graph, and in particular our HCG, consists of points (vertices) and line segments (edges), it is already a topological object in that sense, and thus can contain 1-dimensional holes (loops). But it cannot contain any higher dimensional holes. Moreover, the 1-dimensional holes in the HCG can be generated from different types of hypergraph substructures. Specifically, we define a nest of hyperedges as a sequence of hyperedges, each one a subset of the next. In the case of three hyperedges, this is when hyperedge *e* is completely contained in hyperedge *f*, which is itself completely contained in hyperedge *g*. Then in the HCG, there is a 3-cycle between *e*, *f*, and *g*. These nests are interesting in their own right, but they are easy to identify in the HCG as the completely connected subgraphs; the machinery of homology is not needed to identify nests. Other holes in the HCG encode relationships among the nests and finding those holes can point to more complex behavior patterns. To find these we need to look not at the nests themselves, but at the relationships between them. This is the motivation behind the second step of our process, to enrich the HCG with higher dimensional structure to distinguish between the nests and the relationships among the nests.

For homology to ignore cycles created by these nests and find other complex containment patterns among the nests we must "fill in" the 1-dimensional holes they create. Specifically, we replace every cycle of length 3 in the HCG with a solid triangle, and every set of 4 vertices that are completely connected (known as a *4-clique*, and representing a nested sequence of 4 hyperedges) with a solid tetrahedron. We do this in general for all *k*-cliques (set of *k* vertices that are all pairwise connected) in the HCG, replacing them with *(k-1)*-dimensional analogs of tetrahedra. The result of this procedure is the nesting object.[2] We noted above that homology of the closure identifies "intrinsic" holes in the hypergraph. In the case of the nesting object, we understand that 1-dimensional holes correspond loosely to cycles in which hyperedges are fully contained within intersections of other hyperedges (although the exact structure that is identified through homology of the nesting complex is more nuanced, especially in higher dimensions). We can see this difference between holes in the closure and nesting object illustrated in Figure 6, where the rightmost column shows the nesting object of our three example hypergraphs. What is interesting is that, while the top two hypergraphs had the same closure, it is now the bottom two that have topologically similar nesting objects. Both have a 1-dimensional hole. The cycles are different length, 4 and 6, but since the overall pattern of pairs of intersecting hyperedges containing a hyperedge within their intersection is the same, these two small hypergraph motifs that give rise to the 1-dimensional holes are grouped as topologically similar. The top hypergraph in the figure is missing one of the hyperedges in the intersection of two larger hyperedges which breaks the cycle and so it is not topologically similar through the lens of the nesting object.

We now have all the mathematical background needed to explore the OpTC data through the nesting object of a specific hypergraph construction. We will see how groups of hypergraph motifs that are topologically similar arise from similar behaviors, many of which correspond to adversarial activity.

## Topological signatures in the OpTC data

In the remainder of this article, we delve into examples from the OpTC data where we look for structure in the form of topological signatures. These examples show how the homology of the nesting objects of hypergraphs provided insight into interpretable patterns of activity that were often tied to malicious actions. As previously described, due to the comprehensive nature of the OpTC dataset there are many options for hypergraph constructions. Topological holes that arise from different choices of vertices and hyperedges could lead to the discovery of distinct but equally significant behaviors. For this investigation we focus on hypergraphs where the vertices represent (source IP, host) pairs (as these identify machines) and the hyperedges represent (destination IP, destination port) pairs to model the way that machines interact with other systems on the network. We restrict each hypergraph to a 10-minute time window of data. As a reminder, this means that a vertex is in a hyperedge if there is network flow log with that (source IP, host, destination IP, destination port) tuple during the time window represented by the hypergraph. To simplify exposition, we will refer to the vertices as "machines" rather than their (source IP, host) pair. We will see that significant aspects of the red team activity, including communication with the command and control (C2) server and lateral movement, manifest as nontrivial homology of this hypergraph's nesting object.

---

[2] The technical term is *restricted barycentric subdivision*. If you are interested in learning more about the technical details see [14, 21, 32].

*Adversarial behavior as holes*

Each of the three days of red team activity in the OpTC data had a different APT attack campaign involving different hosts and users and so can be studied independently; here, we focus on the second day. At a high level, the second day represents a custom PowerShell Empire scenario. PowerShell Empire is a toolkit designed to streamline the phases of an APT attack following initial infiltration. It enables attackers to escalate privileges and gather intelligence, while evading detection. For example, by using PowerShell Empire an attacker can run scripts and modules in memory to stay more easily hidden. Specifically, it provides C2 capabilities that facilitate communication between the attacker and the compromised hosts by generating a listener that is protocol based (often HTTP, HTTPS, or DNS), so the communication looks like normal network traffic.

At the beginning of the second day of the red team activity, the users *bantonio* and *rsantilli* were compromised when they opened malicious attachments from phishing emails on their workstations. As soon as they were compromised, *bantonio*'s and *rsantilli*'s machines began to communicate regularly with the two C2 servers, whose IPs we will refer to as C2A and C2B. In the logs, this means that there are records containing *bantonio*'s machine, with the destination IP C2A using the port for HTTPS traffic. Similarly, there are frequent records containing *rsantilli*'s machine, with the destination IP C2B using the HTTP port. In the hypergraphs, this manifests as a hyperedge labeled (C2A, HTTPS) containing a single vertex corresponding to *bantonio*'s machine, and a (C2B, HTTP) hyperedge containing a single vertex corresponding to *rsantilli*'s machine. The former hyperedge appears in all hypergraphs overlapping with the time where *bantonio*'s machine was active, 10:36 until 15:28.[3] The latter hyperedge appears in all hypergraphs overlapping with the time interval 10:40 to 13:25, when *rsantilli*'s machine was active. Later in the afternoon, across different time windows, the single-vertex (C2B, HTTP) hyperedge instead contains vertices corresponding to other machines, consistent with the lateral movement documented in the diary of red team activity.

These singleton hyperedges that correspond to the C2 communications do not, by themselves, create any topological holes in the nesting object. More hyperedges are needed to tie those behaviors together. In addition to the communications with C2A and C2B, *rsantilli*'s and *bantonio*'s machines also communicated with the same domain controller[4] (DC) across two different channels of communication (HTTP and HTTPS) shortly after they were compromised. Consequently, the large, complex hypergraph representing this time interval (Figure 7, left) contains the motif shown in Figure 7 (center). The nesting object of this motif is a cycle of length 4 shown in Figure 7 (right), which is very similar to the bottom example in Figure 6. This example illustrates a dimension 1 hole in the nesting object whose corresponding motif contains hyperedges corresponding to adversary behavior.

---

[3] The port changes from HTTPS to SSH at 13:11 when bantonio starts a reverse ssh connection.
[4] A DC is a machine responsible for managing network security requests. It provides a service to ensure that only users with authorization on a resource are allowed access.

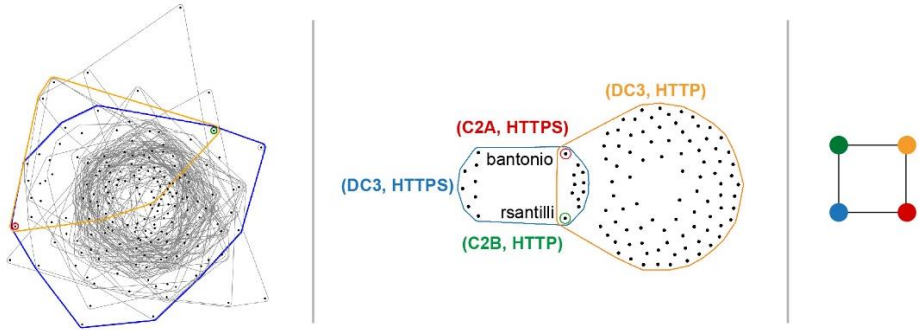

**FIGURE 7.** The full hypergraph corresponding to one 10-minute time window (left) whose nesting object has one 1D hole. The motif representing that hole is shown in the center (hyperedges are also highlighted in the full hypergraph) and nesting object on the right. [**FIGURE 7 ALT TEXT.** A large hypergraph, small subhypergraph, and the nesting object of the subhypergraph.]

    Interestingly, more than half of the motifs we found in hypergraphs within the time period of malicious activity had this motif of two single-vertex hyperedges in the intersection of two large hyperedges (38 out of 64). Figure 8 shows two more motifs (that created holes in the nesting complexes of hypergraphs representing 11:00-11:10 and 13:45-13:55, respectively), that at first glance appear identical to the one in Figure 7, but there are some subtle differences in what the hyperedges represent. In the left motif, the single-vertex hyperedges again contain vertices representing *bantonio*'s and *rsantilli*'s machines, and the blue and orange hyperedges show that these machines were communicating with DCs, but the red hyperedge containing *bantonio*'s machine represents not one destination IP and port hyperedge but 335 of them. Close inspection of the logs shows that these hyperedges represent network flow records that are a downstream effect of the attacker's attempt to elevate the permission level of their account on *bantonio*'s machine using PowerShell Empire. The motif on the right of Figure 8 represents the attacker moving to different hosts using remote desktop sessions. Specifically, the red hyperedge represents when the attacker used a remote desktop session to move to host 974, and the green hyperedge represents when the attacker moved from host 974 to host 005 a few minutes later. These differences illustrate that this common nesting object structure is not just representing one adversary tactic; *rather, this structure is consistent with the way that behaviors interconnect, not with the behaviors themselves.*

    Even when the motifs do not have this structure of two single-vertex hyperedges in the intersection of two large hyperedges, they still often demonstrated connections to adversary behavior. For example, Figure 9 shows a motif that has the same hyperedge containment relationships as the middle hypergraph in Figure 6, and therefore also has a cycle of length 6 as its nesting object. The hyperedges in the motif represent *bantonio*'s and *rsantilli*'s machines talking to their C2 servers (the red hyperedges and green hyperedges, respectively), along with a benign remote desktop session from another machine (the brown hyperedge) and communication with DCs (the three large hyperedges). The commonalities between this and the examples in Figures 7 and 8 illustrate the need for flexibility of motifs and topological signatures, not just in theory but when working with real data. If we had just been searching for motifs that looked exactly like the examples in Figures 7 and 8, we would have missed this.

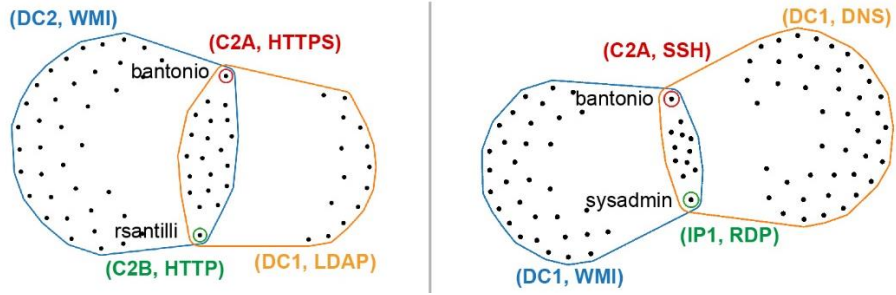

**FIGURE 8.** Two additional subhypergraphs of 10-minute windows that have cycles in their nesting complexes. The structure is the same as in Figure 7 but the behaviors that the hyperedges represent are slightly different. [**FIGURE 8 ALT TEXT.** Two subhypergraphs.]

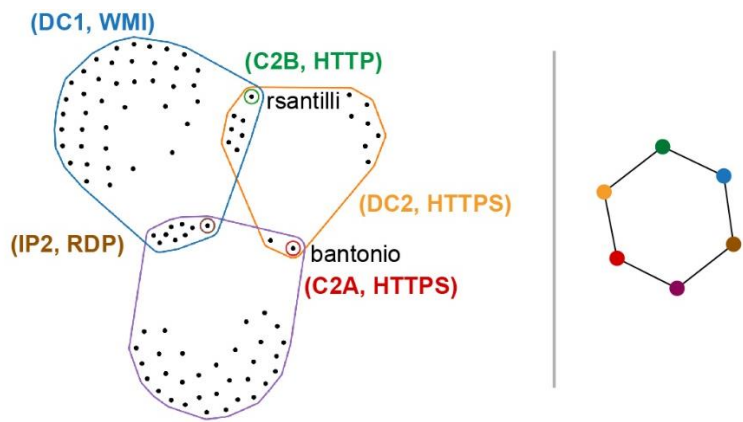

**FIGURE 9.** A subhypergraph of a 10-minute window (left) identified by our algorithm to have a cycle in its nesting object (right). This is a 6-cycle as opposed to a 4-cycle in Figures 7 and 8 but was found as a result of our search for dimension 1 holes. [**FIGURE 9 ALT TEXT.** A subhypergraph and its nesting object.]

## Common interpretations of motifs

In traditional cyber research, signature-based methods search for patterns that match known adversary tactics, techniques, and procedures and raise an alert when a match is found. In contrast, our framework defining topological signatures is designed to discover more flexible patterns that occur in higher dimensional structures of cyber data, and what those patterns mean. But also of great value are methods in exploratory data analysis (EDA), which identify significant patterns from the bottom up. In this exposition we built our examples to clarify the connection to the ground truth adversary behavior. But in practice we made these discoveries through EDA by applying our topology software [29] to the full hypergraph from each time window to detect any instances of homology in its nesting object. We did not know what kinds of behaviors the motifs it found would represent; by inspecting which machines and (destination IP, destination port) pairs the vertices and edges represented, we discovered that many of the motifs were linked to the C2 communication from compromised machines. While not every motif could be linked to malicious activity (although many could, discussed below), a key finding of our work is that most of the motifs were interpretable, in that their hyperedges could be inspected and understood as linked behaviors. Moreover, there were commonalities in the interpretations.

This discovery is summarized in Figure 10, which gives a timeline showing which elements of the red team's activity commonly show up across the motifs. The portion of the timeline where the red team was active is highlighted in red, and the gray cells indicate time windows for which the nesting object of the hypergraph had holes. Note that not all time windows had holes. If the hypergraph did have at least one hole, we show whether any of their corresponding motifs had features we have discussed. The appearance of *bantonio*'s and *rsantilli*'s machines in single-vertex hyperedges are indicated with red and blue dots, respectively; a green dot indicates when the single-vertex hyperedge contained the administrator on host 974; a yellow dot indicates when the single-vertex hyperedge represented an remote desktop protocol (RDP) session (benign or malicious); finally, we indicate when at least one of the large hyperedges of a motif represented communication with a DC with an orange outline. The last two characteristics are not alone indicative of malicious activity; for example, sometimes an RDP session is initiated by an adversary, but other times it is benign. However, we included these characteristics in the figure to emphasize that even when the motif was not linked to malicious activity, it was still interpretable most of the time. In fact, there are only two time windows that have a hole but none of these annotations.

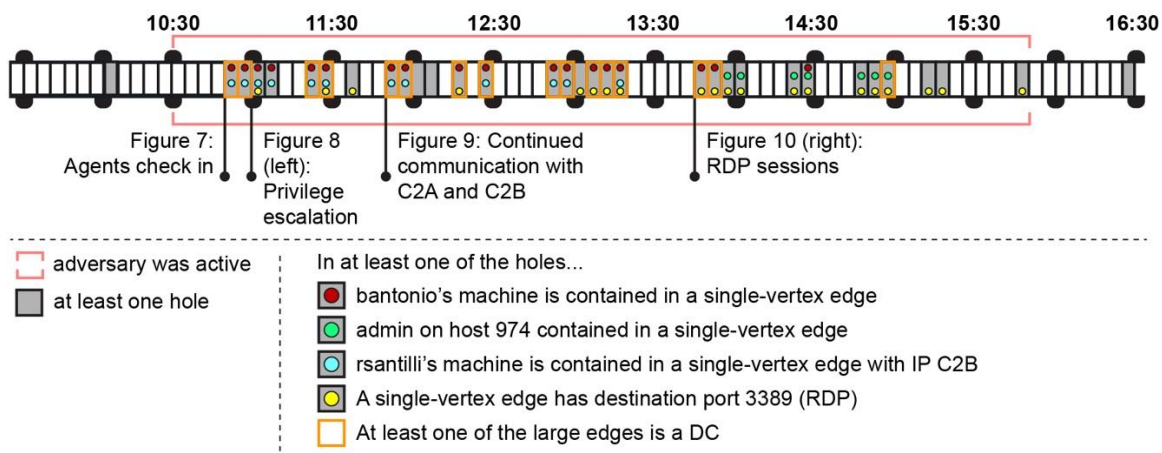

**FIGURE 10.** Summary of findings. We highlight (gray) which 10-minute windows have dimension 1 holes in their nesting objects. We additionally indicate when there is a 1-dimensional hole in a window that contains elements of adversary activity or DC communications. [**FIGURE 10 ALT TEXT.** An annotated timeline.]

It is remarkable that application of our framework—building a hypergraph, creating the nesting object, and computing the homology—discovers behavioral interactions joined together into similar patterns that we can interpret, even though there are differences in the finer details of the activity.

*Evaluating holes as cyber alerts*

Having established the success of our method at discovering and interpreting patterns, we look toward the challenge of turning topological signature identification into a cyber alert specifically designed to capture behavior that aligns with known adversary tactics. Although future work is needed to achieve that goal, in order to identify next steps we describe and study false positives and false negatives as they present in our current analysis.

**False positives.** If critical components of a motif (e.g., the single-vertex hyperedges included in larger hyperedges) could be linked to malicious behavior via process ID in the ground truth, we consider the motif to be malicious, a true positive. Of the 64 motifs in 31 time windows, 47 motifs (73.4%) were malicious. These 47 motifs occurred in 24 of the time windows, and when multiple motifs occurred in one time window, they were often very similar (for example, differing by a single hyperedge). The remainder of the motifs are false positives, but 13 of them were found to be related to benign RDP sessions. This leaves only 4 motifs (6.25%) that were not interpretable.

Another way we could quantify false positives is by counting motifs in hypergraphs constructed over a benign time period in the data with comparable levels of activity to the times of compromise. Unfortunately, the heterogeneity of the OpTC data makes this difficult. To approximate a benign time period we used the same data, but eliminated all network flow logs collected from compromised hosts before recomputing the homology. When we did this, we saw the number of vertices and hyperedges did not change much (Figure 11, left) but the average number of holes per time window decreased from 1.0 to 0.43, and the majority of time windows had no homology (53 out of 64 had no homology, compared to 33 out of 64 with all hosts), which is shown in Figure 11 (right). It may not seem surprising that the number of holes is smaller after removing some vertices and hyperedges. But notice that in Figures 7, 8, and 9 there are many vertices (machines) that we could remove from the motifs without changing the fact that there is a 1-dimensional hole in the nesting object. In these cases, it is the compromised machines (vertices) and the communications resulting from that compromise (hyperedges) that are crucial to the existence of the holes. In other words, it is which vertices and hyperedges we removed that is important.

**False negatives.** Our analysis of false negatives is more nuanced because of the choices built into our framework, namely the choice of hypergraph construction. To mirror the definition of a false positive we could consider any adversary action that was not part of a motif as a false negative. However, there are a variety of different behaviors during an APT attack, some of which may primarily show up in host logs rather than network flow logs. The variety of adversary tactics mean that there will not likely be a single hypergraph construction that detects all malicious activity. While holes in the nesting object of the hypergraph construction we focused on seem to consistently identify similar patterns of behavior, more work is needed to discover hypergraph constructions that can be used to detect other adversarial behaviors in this dataset and others.

Additionally, we point out that when the adversary was active, the single-vertex (C2A, HTTPS) and (C2B, HTTP) hyperedges were present in each hypergraph. But in some time windows these hyperedges are not part of a motif giving rise to a hole. This is because our framework is designed to detect the linkage of behaviors in interesting ways, rather than single behaviors. Once adversary behavior is discovered in one time period as part of an instance of a topological signature, the individual components of the signature can be used to seed analyst queries for other instances of those behaviors.

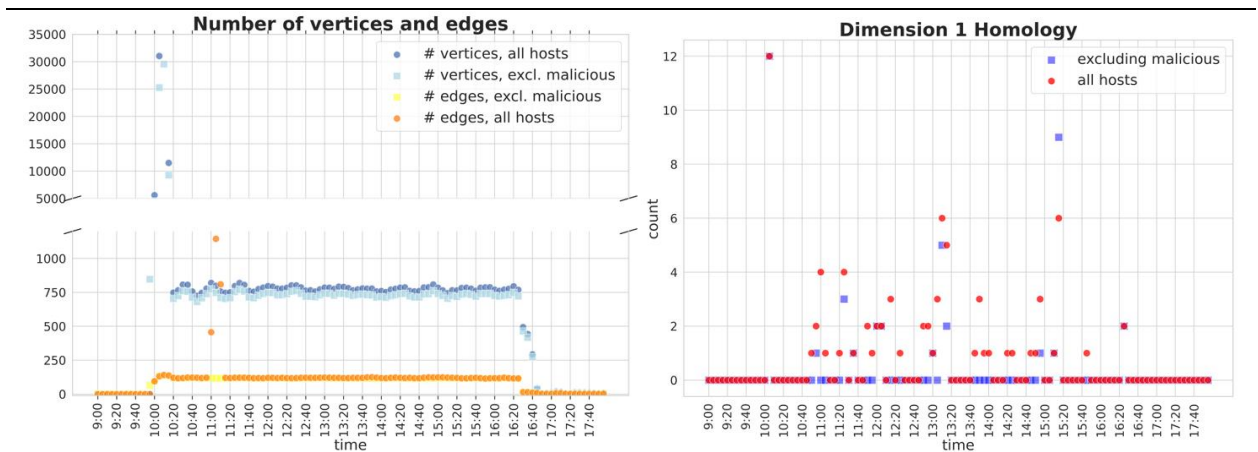

**FIGURE 11.** The number of vertices and hyperedges in hypergraphs with (source IP, host) vertices and (destination IP, destination port) hyperedges constructed from Day 2 of the OpTC data (left) and the number of dimension 1 holes in these hypergraphs (right). To quantify homology unrelated to adversarial activity we constructed hypergraphs both using data from all the hosts and a subset of the data that excludes the compromised hosts. Note that malicious activity occurs from 10:28 to 15:42. [**FIGURE 11 ALT TEXT.** A scatter plot showing the number of vertices and hyperedges and a scatter plot showing the number of dimension one holes.]

## Conclusion

The problem of developing new cyber tools that enable effective attack detection in a way that keeps up with adversary developments is challenging, in part because of the unknown complexity within the data. The framework we introduce here, identifying behavior linkage patterns through topological signatures, answers the challenge of stepping out of Flatland into the unknown dimension of Cyberland and points to a wide area of potential research. Our exploration of the OpTC data provides one example of its utility to make sense of this complex cyber data. Unlike traditional signature-based cyber tools, our methods did not set out to define patterns based on known adversary techniques, rather we modeled the complexity using hypergraphs and looked for structure via homology. It was very surprising and exciting that homology of the nesting objects took these hypergraphs with around 750 vertices and 100 hyperedges each and found small hypergraph motifs that were typically interpretable and often included adversary activity. There is still a need for more theory to understand the homology of topological translations of hypergraphs, interpretation to flesh out how topological signatures relate to user and adversary behaviors, and a bridge to be built between theoreticians and cyber analysts to ensure both sides understand and can help shape the perspective of the other. Our team will continue to explore these directions and we look forward to collaborations and partnerships as we continue our journey through Cyberland.